# The role of vacancies, impurities and crystal structure in the magnetic properties of $TiO_2$.

Mariana Weissmann[a] [*] and Leonardo A. Errico[b]

[a] Departamento de Física, Comisión Nacional de Energía Atómica, Av. del Libertador 8250, 1429 Buenos Aires, Argentina.

[b] Departamento de Física - Instituto de Física La Plata (CONICET), Facultad de Ciencias Exactas, Universidad Nacional de La Plata,

C.C. 67, 1900 La Plata, Argentina.

---

## Abstract

We present an *ab initio* study of pure and doped $TiO_2$ in the rutile and anatase phases. The main purpose of this work is to determine the role played by different defects and different crystal structures in the appearance of magnetic order. The calculations were performed for varying impurity and vacancy concentrations in both $TiO_2$ structures. For Co impurities the local magnetic moment remained almost independent of the concentration and distribution while for Cu this is not the case, there is magnetism for low concentrations that disappears for the higher ones. Impurity-impurity interactions in both structures favor linear ordering of them. Magnetism in un-doped samples appears for certain vacancy concentrations and structural strain. © 2001 Elsevier Science. All rights reserved

---

## 1. Introduction

Diluted magnetic semiconductors and insulators, especially those having room-temperature (RT) ferromagnetism, have attracted great interest in the last years due to their potential applications in spintronic devices [1]. Many systems, such as GaN:Mn [2], GaN:Cr [3], ZnO:Co [4], ZnO:Mn [5], $TiO_2$:Co [6, 7], or $CdGeP_2$:Mn [8], have been reported to have RT-ferromagnetism. In particular, the first report of robust RT-ferromagnetism by Matsumoto *et al.* [6] generated a tremendous interest in Co-doped $TiO_2$ films. However, a large spread of Curie temperatures and magnetizations were reported since then (see, for example, [9]). The growth conditions, the substrate structure and temperature, the amount of oxygen present, all seem to influence the ferromagnetic behaviour. Theoretical and experimental studies have been intensively performed [10], but a consensus on the origin of the experimentally observed ferromagnetic coupling has not yet been reached. The use of these systems in spintronic devices has today two types of limitations: the first one is the preparation of a homogeneous solid solution of the dopant in the oxide host, avoiding phase segregation. The second one is understanding the mechanism associated with the ferromagnetic ordering [10, 11, 12].

In the present work we present a computational *ab initio* study of un-doped rutile and anatase $TiO_2$ with oxygen vacancies and of $TiO_2$ doped with a magnetic ion (Co) and a non-magnetic one (Cu). We put special care in the structural relaxations and in the effects of impurity concentration. We have not studied here cases with both impurities and vacancies, as in that situation

[*] Corresponding author. Tel.: +0-000-000-0000 ; fax: +0-000-000-0000 ; e-mail: author@institute.xxx .

there is charge compensation and possibly few charge carriers.

## 2. Method of calculation

The spin-polarized electronic-structure calculations presented in this work were performed with two different *ab initio* codes, Wien2K [13] and Siesta [14]. The Wien2K is an implementation of the full-potential linearized-augmented-plane-wave method (FP-LAPW [15]). Exchange and correlation effects are treated within density-functional theory using the local spin density approximations (LSDA [16]). The atomic spheres radii ($R_i$), used for Ti and O were 1.01 and 0.85 Å, while for Co and Cu we used $R_i$=1.06 Å. The parameter $R_{KMAX} = R_{mt} * K_{MAX}$, which controls the size of the basis-set in these calculations, was chosen as 7 ($R_{mt}$ is the smallest muffin tin radius and $K_{MAX}$ the largest wave number of the basis set). We introduced local orbitals to include Ti-3$s$ and 3$p$, O-2$s$ and Co- and Cu-3$p$ orbitals. The number of $k$-points was increased until convergence was reached.

The SIESTA code uses a linear combination of numerical real-space atomic orbitals as basis set and norm-conserving pseudopotentials [17]. Although both contain the LSDA approximation, the Siesta code is much faster for relaxation of the structures and gives the results in terms of atomic orbitals, so that one can separate the contribution of each orbital to the magnetic moment. The disadvantage is that pseudopotentials and basis sets have to be proved adequate for each calculation. On the contrary, the Wien2K code can be converged in the number of plane waves and the number of $k$-points, leaving no parameters free. But the disadvantage is that the charge in the interstitial region can not be separated in the different symmetries. Our usual strategy was to relax the structures with the Siesta code and then use those atomic positions as input for the Wien2K calculation. The total energy was always lower than that of the un-relaxed cases.

## 3. Results and discussions

Titanium dioxide crystallizes in three different structures: anatase (tetragonal, *I4₁/amd*, $a=b$=3.782 Å, $c$=9.502 Å), rutile (tetragonal, *P4₂/mnm*, $a=b$=4.5854 Å, $c$=2.9533 Å), and brookite (rhombohedral, *Pbca* $a=b$=5.436 Å, $c$=9.166 Å) [18]. However, only rutile and anatase have applications and are of interest here. Their unit cells are shown in Fig. 1. In both structures, the basic "building bricks" consist of a titanium atom surrounded by six oxygen atoms in a more or less distorted octahedral configuration. In both structures, the two bonds between the titanium and the oxygen atoms at the apices of the octahedron are slightly longer.

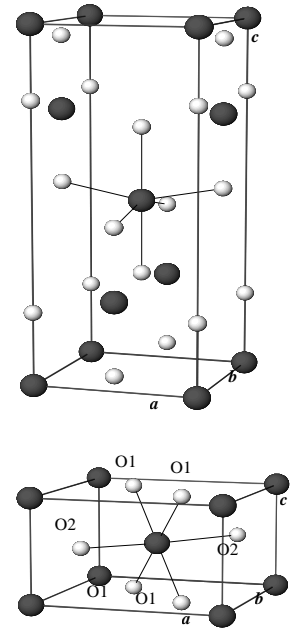

Fig. 1: Bulk structures of anatase (up) and rutile (down) TiO₂. Black and white balls represent Ti and O, respectively.

### a) Effect of vacancies in un-doped TiO₂.

Different experimental and theoretical works suggest that oxygen vacancies play an important role in the origin of ferromagnetism of doped TiO₂ [10]. In addition, ferromagnetism was discovered recently in thin films of pure TiO₂ [19] and other nonmagnetic oxides [19, 20]. One of the mechanisms proposed in order to explain the observed ferromagnetism is based in the presence of intrinsic oxygen vacancies, which act as donors, leading to *n*-type doping of the material. In order to analyze the effect of oxygen vacancies in doped $TiO_2$ we first studied the reduced pure system, $TiO_{2-\delta}$, taking into account the structural distortions produced by the vacancies using the super-cell (SC) method. We employed a small SC consisting of two unit cells of the rutile structure stacked along the *c*-axis and we removed an oxygen atom ($Ti_4O_7$) from this SC.

$TiO_2$ is an ionic crystal, therefore an oxygen vacancy produces a large structural relaxation due to the repulsion of nearby titanium ions, and leaves a big hole in its place. In effect, the cations nearest neighbors to the vacancy site are repelled from it. To study this relaxation we used both the Wien2K code, with the mini routine, and the Siesta code with the conjugate gradient procedure. In the first case the symmetry around the vacant site is maintained on relaxing while in the second case it is not. The final atomic positions are different: in one case the vacancy-site distances to its Ti nearest neighbors are 2.16 and 2.34 Å; in the other they are 1.97 and 2.40 Å (all are larger than the oxygen-Ti nearest neighbor distances in stoichiometric $TiO_2$ of 1.94 and 1.98 Å). The energies of the two relaxed unit cells with one vacancy differ by less than $10^{-4}$ Ryd, but although they were both calculated with the Wien2K code up to convergence one of them leads to zero magnetic moment while the other has 1.6 $\mu_B$/SC. This surprising result is obtained when the cell size is kept fixed at the experimental value and suggests that strain may be the reason for obtaining ferromagnetism in un-doped oxide films. This strain could be due to the substrate, to the fast growth in an oxygen poor atmosphere or to interstitial atoms nearby. Calculations with a larger unit cell and only one vacancy, both in rutile and anatase structures, were performed but there was no magnetic moment. We conclude that the appearance of a magnetic moment depends on the vacancy concentration and on the strain of the structure.

**b) Effect of concentration and distribution of impurities.**

For this study we restricted ourselves to one magnetic (Co) and one non-magnetic impurity (Cu) with no oxygen vacancies. In the case of the rutile phase, the first SC considered consisted of two unit cells of $TiO_2$ stacked along the *c*-axis with one Ti atom replaced by the impurity. The resulting 12-atoms SC ($Ti_3RO_8$) has dimensions $a'=b'=a$, $c'=2c$. The impurity concentration in this SC is 25%. The second SC employed consisted of 8 units cells of $TiO_2$, corresponding to an impurity concentration of 6.25%. The resulting 48-atoms SC ($Ti_{15}RO_{32}$) has dimensions $a'=b'=2a$, $c'=2c$. In the case of the anatase phase we performed calculations with the simple 12-atoms unit cell and also with a 48-atoms SC ($Ti_{15}RO_{32}$, $a'=b'=2a$, $c'=c$).

We found that the presence of Co impurities induces local geometrical distortions (contractions) in the first oxygen neighbors ($O_{NN}$) of the impurities. The bond lengths Co-$O_{NN}$ are reduced by about 0.05 Å. In the case of Cu the structural distortions induced by the impurity are anisotropic, the Cu-O1 bond lengths are shorter while the Cu-O2 distances are enlarged by about 0.04 Å.

|  | Rutile | Anatase |
|---|---|---|
|  | $\mu^{imp} / \mu^{SC}$ ($\mu_B$) | $\mu^{imp} / \mu^{SC}$ ($\mu_B$) |
| $TiO_2$:Co, 12-atoms SC | 0.63 / 0.98 | 0.73 / 1.01 |
| $TiO_2$:Co, 48-atoms SC | 0.64 / 1.14 | 0.73 / 0.96 |
| $TiO_2$:Cu, 12-atoms SC | 0.00 / 0.00 | 0.00 / 0.00 |
| $TiO_2$:Cu, 48-atoms SC | 0.47 / 1.01 | 0.37 / 0.97 |

Table 1: Magnetic moments in the muffin tin sphere of the impurities ($\mu^{imp}$) and in the super-cell ($\mu^{SC}$) for the two concentrations and structures studied.

Results for the magnetic moments are summarized in Table 1. We see that the same magnetic moment of 1.0 $\mu_B$/SC was found for Co in both structures and concentrations, but that for Cu there is no magnetic moment in the small unit cells while it is 1.0 $\mu_B$/SC in the larger systems. The appearance of magnetism in Cu-doped $TiO_2$ agrees with experimental results [21]. The same dependence of the magnetic moment on concentration of Cu was also found in calculations for ZnO [22]. This concentration dependence shows again, as for the un-doped system, that the variety of results obtained experimentally for different samples also appears in *ab initio* calculations, even though these are performed for bulk materials and at 0K.

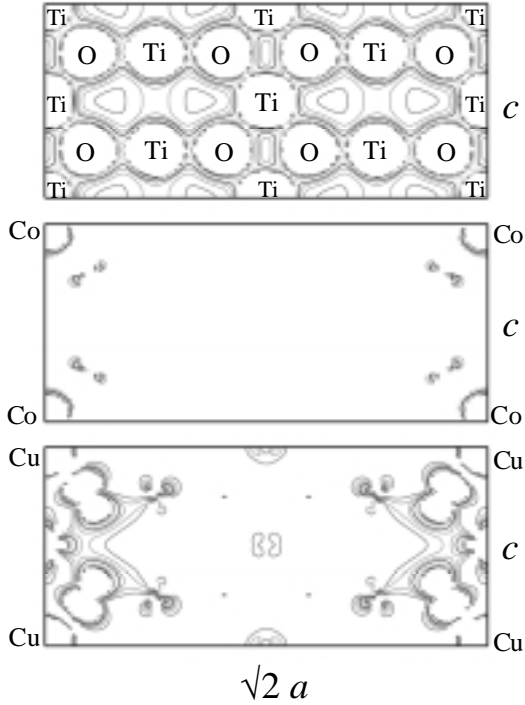

Fig.2: (up) Charge density for the (110) plane of the rutile structure of TiO$_2$ in the 48-atoms SC; (middle) Spin density for the same plane with one Co impurity per SC; (down) Spin density for the same plane with one Cu impurity per SC.

Table 1 also shows that Co spin-polarization occurs mainly at the impurity sites while the O$_{NN}$ are polarized only up to 0.06 $\mu_B$/SC, but that in the case of Cu the magnetic moment is delocalized. In Fig.2 we present the charge and spin densities for the rutile structure in the (110) plane, where this difference in spin localization is clearly evident. In Fig. 3 we present the charge and spin densities for the (001) plane in both structures. As can be seen, the hybridization is also very different for the two different impurities and this, together with the different spin localization raises the question if the mechanism for long range ferromagnetic order could be the same for Co and Cu.

The densities of states (DOS) of these systems are shown in Fig. 4. In all cases, the impurity states (*d* states) are mainly located in the energy-gap of TiO$_2$ and the Ti and O states are not much affected by the doping. The features of the 48-atoms SCs are also valid for the 12-atoms SCs, the only differences being the larger values of the DOS at the Fermi levels and the increased structures from impurity-induced states. These results are in good agreement with previous calculations, for example those reported by Janisch *et al.* [23] for the case of Co in anatase TiO$_2$.

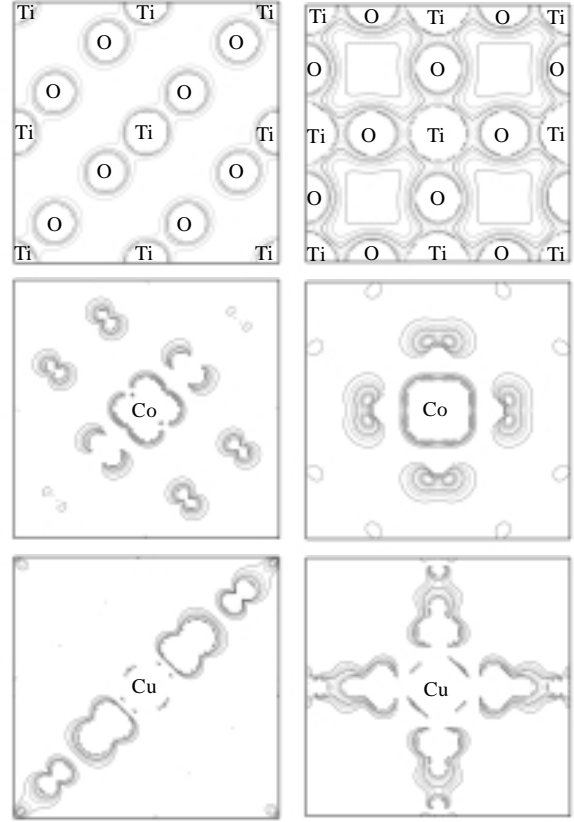

Fig.3: Charge (up) and spin densities (middle and down) in the (001) plane of rutile (left) and anatase (right) for Co and Cu doped TiO$_2$. There is one impurity in the 48 atom SC.

Next we substituted a second Ti atom by an impurity in the 48-atom SC, leading to an impurity concentration of 12.5 %, and allowing us to compare the energies of ferro- and antiferromagnetic orderings. We varied the positions of the impurity atoms to study the preferred configurations and the dependence of the magnetic moment on the impurity distribution. We found that in all cases the lowest energy corresponded to a ferromagnetic alignment, as was found in Ref [22] for ZnO.

Our results for the rutile structure show that if the Cu atoms are located at the shortest possible distance to each other the system has its lowest energy and highest magnetic moment. This corresponds to impurities located along the *c*-axis of the rutile structure, where they are at 2.90 Å and form a row of impurities due to the periodicity. It would be very interesting to build experimentally these linear structures and to find out if they validate these results. It is important to mention that when Cu impurities are considered, there is an increase of the magnetic moment from 1.1 $\mu_B$/SC to 3.2 $\mu_B$/SC (see Table 2). In the case of Co impurities, the lowest energy distribution also corresponds to Co at the shortest possible distance, but the magnetic moment is nearly independent of the Co distribution in the host. The magnetic moment in these cases is 2.0 $\mu_B$/SC, two times the magnetic moment corresponding to the case of one Co impurity (see Table 2). This difference could be due to the greater localization of spin in the case of Co impurities.

For the anatase structure the nearest neighbor positions do not form a row and the lowest energy for the pair of impurities is when they are second neighbors, forming a row along the *a*-axis at a distance of 3.78 Å from each other. In the case of Cu there is a decrease of the magnetic moment, to 0.80 $\mu_B$/SC, while for Co we obtained similar results to those found for the rutile structure.

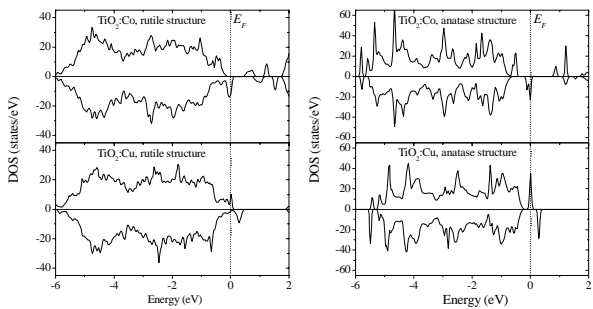

Fig.4: Densities of states of Co- and Cu-doped rutile and anatase TiO₂. Energies are referred to the Fermi level ($E_F$). There is one impurity in the 48-atoms SC.

## 4. Conclusions

From these calculations we conclude that magnetism in un-doped transition metal oxide films is possibly due to vacancies and structural strain, produced either by the substrate or by the growth procedure. Slightly different relaxations with a fixed cell size, calculated with the same code and precision and energetically degenerate, give very different magnetic moments.

For the doped system, we analyzed the case of a magnetic impurity (Co) and a non-magnetic one (Cu). Co impurities have a localized magnetic structure, with 1.0 $\mu_B$/SC and a very small interaction among them. On the other hand, Cu impurities produce a magnetic moment for very small concentrations only, and it is delocalized so that about 0.5 $\mu_B$ can be attributed to Cu and the rest to neighbor atoms.

Concerning the impurity distribution in the host lattices, the impurities prefer to be located in rows, along the *c*-axis of the rutile structure and the *a*-axis of anatase. The magnetic moment does not depend on the distance between Co atoms but it depends strongly for Cu. It would be very interesting to produce rows of impurities experimentally. The differences in the spin densities obtained between Co and Cu doping, and between rutile and anatase structures, raise the question if the mechanism for long range ferromagnetic order is the same for Co and Cu doping, and also if it is the same for undoped films.

## Acknowledgements.

This work was partially supported by ANPCyT under PICT98 03-03727 and 03-10698 , by CONICET (under PIP006/98, PEI 6174, PIP6032 and PIP 6016), Fund. Antorchas, Argentina, and TWAS, Italy, RGA 97-057. M.W. and L.A.E are members of Carrera del Investigador, CONICET, Argentina. We are indebted to Dr. Javier Guevara for critical reading of the manuscript and helpful suggestions.

## References.

| | **Rutile** | | **Anatase** | |
|---|---|---|---|---|
| | $\mu^{SC}$ ($\mu_B$) | $d$ (Å) | $\mu^{SC}$ ($\mu_B$) | $d$ (Å) |
| **TiO$_2$:Cu, 1 Cu-impurity** | 1.01 | - | 0.97 | - |
| **TiO$_2$:Cu, 2 Cu-impurities** | 3.2 | 2.96 | 0.8 | 3.78 |
| **TiO$_2$:Cu, 2 Cu-impurities** | 1.8 | 5.90 | 2.0 | 5.46 |
| | | | | |
| **TiO$_2$:Co, 1 Co-impurity** | 1.14 | - | 0.96 | - |
| **TiO$_2$:Co, 2 Co-impurities** | 2.0 | 2.96 | 2.0 | 3.78 |
| **TiO$_2$:Co, 2 Co-impurities** | 2.0 | 5.90 | 2.0 | 5.46 |

Table 2: Magnetic moments per SC ($\mu^{SC}$) obtained for the systems Ti$_{15}$R$_1$O$_{32}$ and Ti$_{14}$R$_2$O$_{32}$ (R: Co and Cu). Of all the possible configurations for the 2 impurities we only show the one with lowest energy and the one where the impurities are farther apart. We denote with $d$ the impurity-impurity distance.